\documentclass{appolb}
\usepackage{amsmath}
\usepackage{amsfonts}
\usepackage{amssymb}
\usepackage{epsfig}

\begin{document}
\title{$S-P$ wave interference in the $K^+K^-$ photoproduction on hydrogen
\thanks{Presented at Cracow School of Theoretical Physics, XLV Course, \\
Zakopane, June 7, 2005}%
}
\author{\L{}ukasz Bibrzycki, Leonard Le\'sniak
\address{H. Niewodnicza\'nski Institute of Nuclear Physics PAN, PL 31-342 Krak\'ow, Poland}
\and
Adam P. Szczepaniak
\address{Physics Department and Nuclear Theory Center,\\
Indiana University, Bloomington, IN 47405, USA}
}
\maketitle
\begin{abstract}
We have studied the partial wave interference effects to obtain a new information about the contribution of the $S-$wave to the cross section of the $K^+K^-$ 
photoproduction on hydrogen. The $K^+K^-$ photoproduction channel for the effective masses around 1 GeV is dominated by the $\phi(1020)$ resonance with only a small fraction of
events coming from decays of scalar resonances $f_0(980)$ and $a_0(980)$. However, a careful analysis of
angular distributions of the outgoing kaons shows that the $S-$ wave adds an asymmetry to the
angular distribution of kaons. A fairly precise estimation of the $K^+K^-$ photoproduction cross section in the $S-$ wave has been obtained.
\end{abstract}
\PACS{13.60.Le, 25.20.Lj}
  
\section{Introduction}
Both experimental and theoretical analyses of the near threshold photoproduction of the $K^+K^-$ pairs are
crucial for a better understanding of the nature of scalar mesons $f_0(980)$ and $a_0(980)$. Moreover
there exists a hypothesis that the $f_0(980)$ may be a $K\overline{K}$ bound state. The interest in the near
threshold $K^+K^-$ production dynamics and a relatively  large coupling of the photon to vector mesons encouraged
experimentalists to perform a series of experiments in seventies and eighties. Our investigations base on
the results obtained by Behrend et al. \cite{Behrend} at DESY and Barber et al. \cite{Barber} at
the Daresbury Laboratory. These experiments showed unequivocally that the $S-$ wave participates in the $K^+K^-$
photoproduction on hydrogen which can be seen in figures showing the moments of angular distribution
as a function of the $K^+K^-$ effective mass $M_{KK}$. However, in these early investigations the number of independent amplitudes taken into consideration was limited to three. 
This model limitation
combined with large experimental uncertainties resulted in very big differences between the total cross
sections reported by two experiments. The value of the total $K^+K^-$ photoproduction cross section ranged
from (2.7$\pm$1.5) nb derived from the data of Behrend et al. to (96.2$\pm$20) nb corresponding to the data of Barber et al. Contrary to previous
experimental analyses we
include all the independent amplitudes i.e. 4 amplitudes in the $S-$ wave and 12 amplitudes in the $P-$
wave. Moreover our approach takes into account all 6 moments of angular distribution (including the moment 
$\langle Y^0_{\; 0}\rangle$ proportional to the effective
mass distribution) which can be constructed from the spin 0 and spin 1 amplitudes. In the experimental analyses only two moments $\langle Y^0_{\; 0}\rangle$ and $\langle Y^1_{\; 0}\rangle$ were fitted. The
data provided by two experiments correspond to two slightly different kinematic regions defined below:
\begin{enumerate}
\item $E_\gamma$=4 GeV, $-t<1.5$ GeV$^2$, 0.997 GeV$<M_{KK}<$1.042 GeV \cite{Barber},
\item $E_\gamma$=5.65 GeV, $-t<0.2$ GeV$^2$, 1.005 GeV$<M_{KK}<$1.045 GeV \cite{Behrend}.
\end{enumerate}
Apart from analysing accessible data we have also applied the constructed model to the case of the incident photon energy $E_\gamma$=8 GeV,
corresponding to the value designed for the future energy upgraded  facility at JLab \cite{Dzierba}.

\section{Description of the model}
Here we present very briefly only the most important ingredients of our model referring the reader to our
previous papers \cite{Ji,acta,BLS} for a more extensive reading. The starting point of our
investigation was the construction of the partial wave decomposed amplitudes for the $K^+K^-$ photoproduction process.
The amplitudes are defined by
\begin{equation} \label{ampl_gen}
T_{\lambda_\gamma \lambda \lambda'}(E_\gamma,t,M_{KK},\Omega)=\sum_{L=0,1;M}
T^{L}_{\lambda_\gamma \lambda \lambda';M}(E_\gamma,t,M_{KK})Y^L_{\; M}(\Omega),
\end{equation}
where
L and M denote the angular momentum of the $K^+K^-$ subsystem and its projection on the helicity axis,
$\lambda_\gamma$, $\lambda$ and $\lambda'$ are the helicities of the photon, the incoming proton and the outgoing proton, respectively,
$t$ is the momentum transfer squared and
$\Omega=(\theta,\phi)$ denotes the solid angle of the outgoing $K^+$.
The angles and momenta are defined in the so called s-channel helicity frame. This frame coincides with the $K^+K^-$ rest frame in which the z-axis is directed opposite to the recoil proton momentum and the y-axis is perpendicular to the $\phi p$ production plane. 
\begin{figure}[h] \label{born_graphs}
  \begin{center}
  \epsfig{file=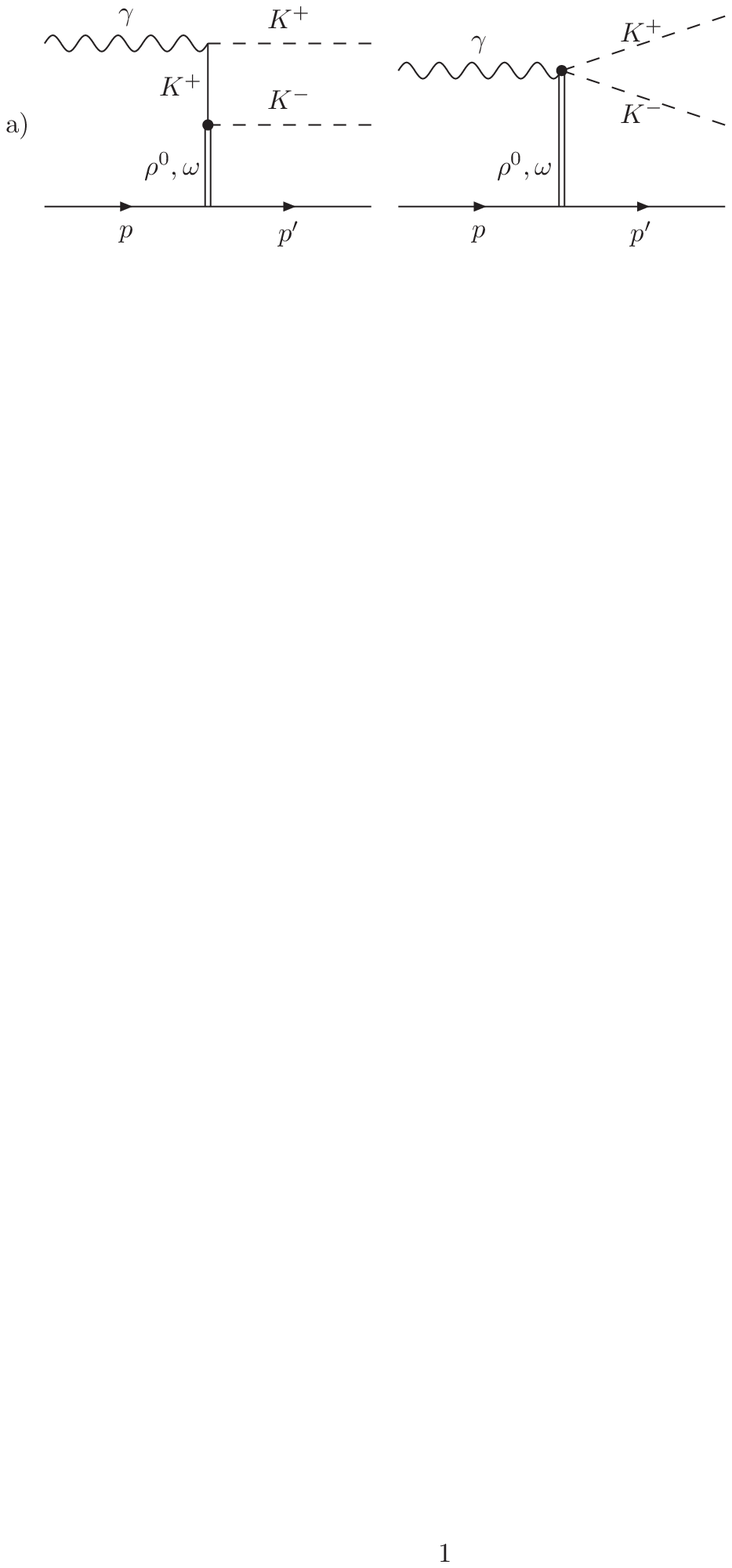, scale=.8}
  \epsfig{file=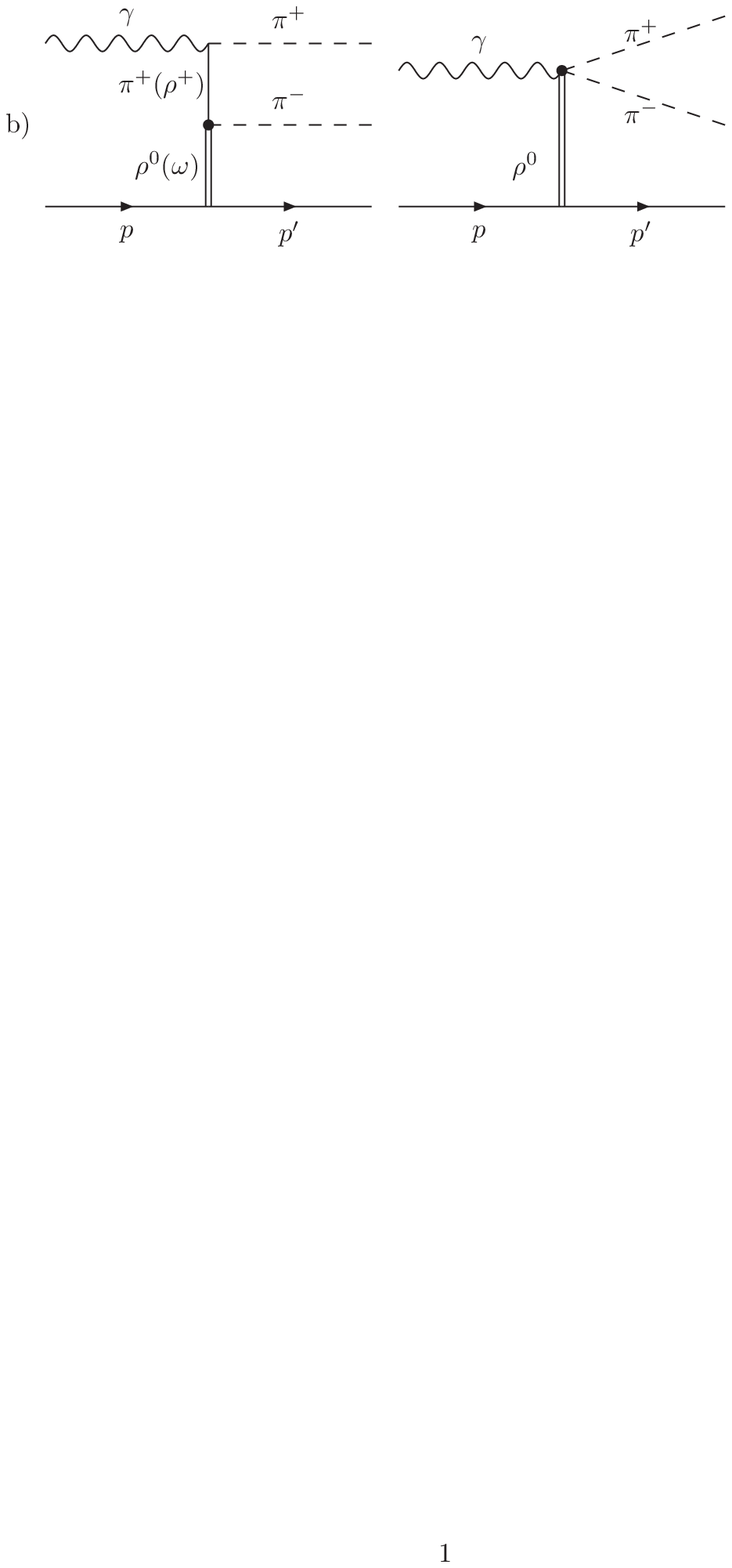, scale=.8}
  \caption{Some diagrams representing the $K^+K^-$ (a) and the $\pi^+\pi^-$ 
  (b) Born photoproduction amplitudes}
  \end{center}
\end{figure}
\subsection{$S-$wave amplitude}
In our model the $S-$ wave component of the $K^+K^-$ photoproduction amplitude is parameterized by 
the $t-$ channel exchange of the $\rho$  and
$\omega$ vector mesons. The amplitude has been decomposed into the isoscalar part and the
isovector part in the following way: 
\begin{equation} \label{isos_dec}
A^S(I)=\frac{1}{2}[A^S(I=0)+A^S(I=1)].
\end{equation}
Additionally the amplitude has been factorised into the Born factor $A^B_j(I)$ and the factor $t_{jf}(I)$ responsible for
the final state interactions according to the formula:
\begin{equation} \label{born_fsi}
A^S(I)=\sum_{j=\pi\pi,K\overline{K}}A^B_j(I)t_{jf}(I).
\end{equation}
The Feynman graphs which contribute to the $S-$ wave Born amplitudes are schematically
shown in \mbox{Fig. 1.} 
The final state interaction factor $t_{jf}(I)$ is of the form $t_{jf}(I)\sim\frac{1}{2}[\delta_{jf}+S_{jf}(I)]$. This factor accounts for the $\pi^+\pi^-$ and
$\pi^0\pi^0$ intermediate states, and for the $K^+K^-$ elastic rescattering.
The diagrams describing the $S-$ wave
amplitudes are schematically drawn in Fig. 2.
The isoscalar
$S-$ matrix parameterised in terms of the channel phase shifts $\delta$ and inelasticity $\eta$ reads:
\begin{equation}
S(I=0)=
\left(\begin{array}{cc}
\eta e^{2i\delta^{I=0}_{\pi\pi}} &i\sqrt{1-\eta^2}e^{i(\delta^{I=0}_{\pi\pi}+
\delta^{I=0}_{K\overline{K}})}\\
i\sqrt{1-\eta^2}e^{i(\delta^{I=0}_{\pi\pi}+
\delta^{I=0}_{K\overline{K}})}& \eta e^{2i\delta^{I=0}_{K\overline{K}}}
\end{array}\right).
\end{equation}
The isovector $S-$ matrix is defined analogously.

We use two kinds of propagators to describe the propagation of the $\rho$ and $\omega$ mesons
in the Born diagrams: the normal propagator $1/(t-m^2)$ or the Regge-type propagator
\begin{equation}
-[1-e^{-i\pi \alpha(t)}]\Gamma(1-\alpha(t))(\alpha's)^{\alpha(t)}/(2s^{\alpha_0}),
\end{equation}
where $m$ is the mass of the exchanged vector meson and 
$\alpha(t)=\alpha_0+\alpha^{'}(t-m^2)$ denotes the Regge trajectory of the vector meson in which $\alpha_0$=1 and $\alpha^{'}$=0.9 GeV$^{-2}$.
\begin{figure} \label{full_graphs}
\begin{tabular}{cc}
\includegraphics[width=.45\textwidth]{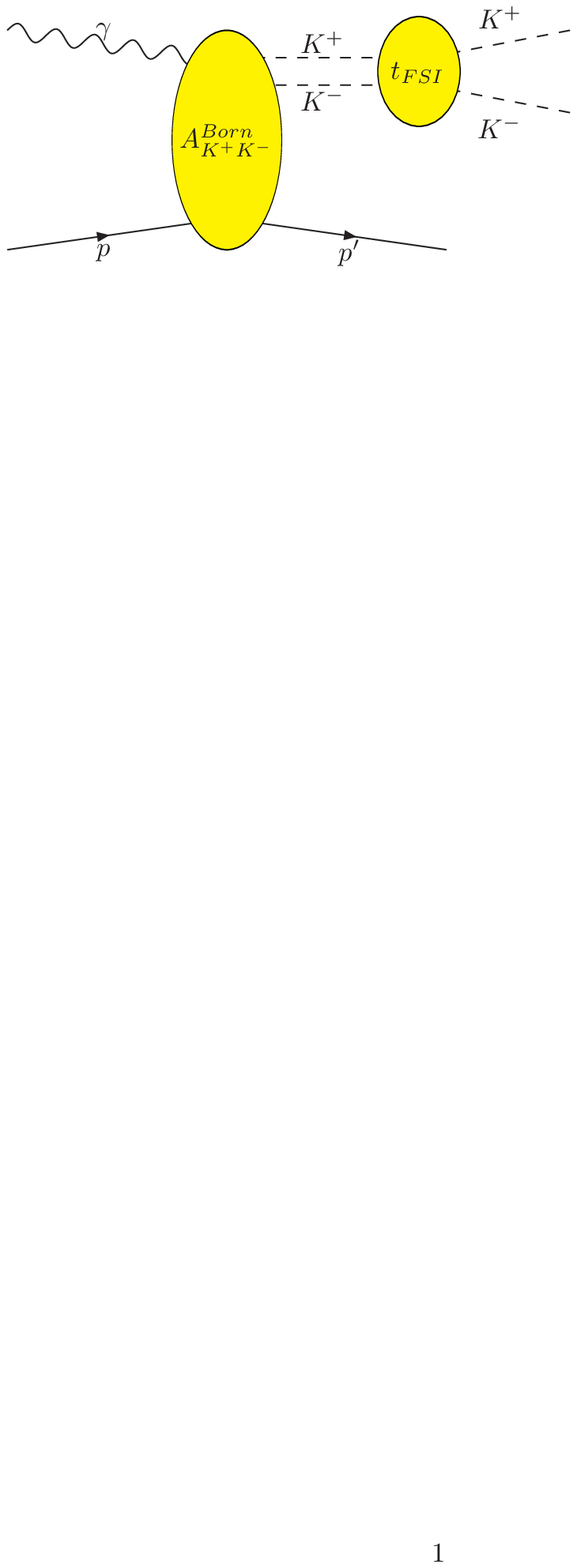}&
\includegraphics[width=.45\textwidth]{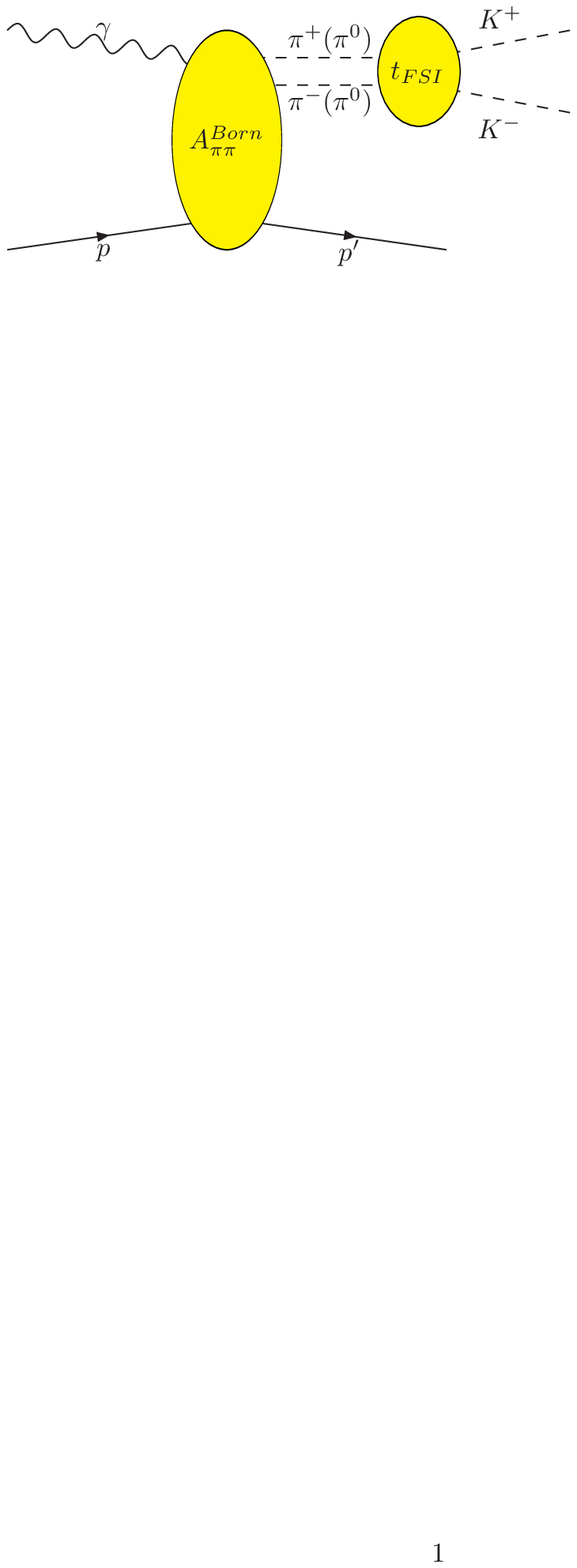}\\ 
\end{tabular}
\caption{Diagrams for elastic $K^+K^-$ rescattering and inelastic $\pi \pi \rightarrow K^+K^-$
transition}
\end{figure} 
\subsection{$P-$ wave amplitude}
We have assumed the pomeron exchange as a dominant
reaction mechanism of the $K^+K^-$ photoproduction in the $P-$ wave . This approach is strongly supported by the OZI rule and previous experimental results. The
Feynman diagram for the $P-$ wave amplitude is shown in Fig. 3.
\begin{figure} \label{feyn_p}
\begin{center}
\includegraphics[width=.55\textwidth]{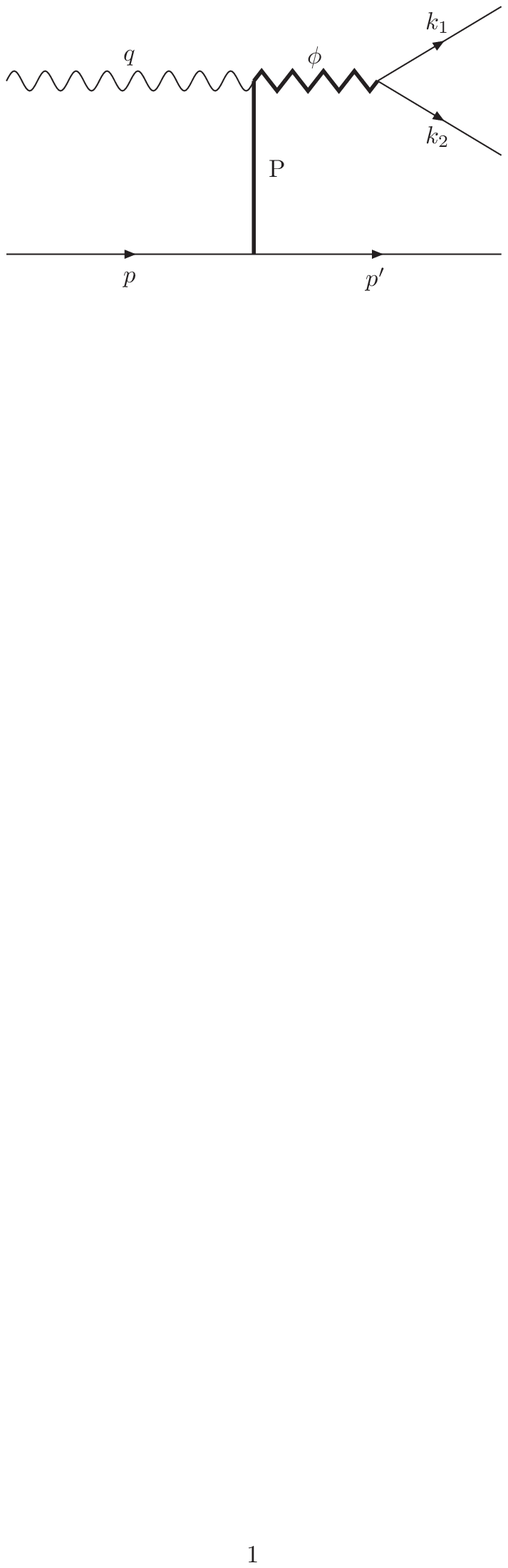}
\caption{Feynman diagram for the $P-$ wave $K^+K^-$ photoproduction on hydrogen}
\end{center}
\end{figure}
The general form of the $P-$ wave amplitude is
\begin{equation} \label{ampl_p}
A^P_{\lambda_\gamma,\lambda,\lambda',M}=\overline{u}(p',\lambda')J^P_
{\mu M} \varepsilon^{\mu}(q,\lambda_\gamma)u(p,\lambda),
\end{equation}
where $q$ is the four-momentum of the incident photon, $\varepsilon^{\mu}$ is the polarisation vector of the photon,
$p$ and $p^{'}$ are the four-momenta of the incoming and recoil proton, and the current $J^P_
{\mu M}$ is defined as follows:
\begin{equation}
J^{P}_\mu=\frac{iF(t)}{M_{\phi}^2-M_{KK}^2-iM_{\phi}\Gamma_\phi}[\gamma^\nu
q_\nu(k_1-k_2)_\mu-q^\nu(k_1-k_2)_\nu\gamma_\mu].
\end{equation}
The $M_\phi$ and $\Gamma_\phi$ denote the mass and width of the $\phi$ resonance, and $k_1$ and $k_2$ are the $K^+$ and $K^-$ four-momenta.
The function $F(t)$ is suitably parameterised to reproduce the experimental differencial cross section $d\sigma/dt$ for the photon energies of 4 GeV or \mbox{5.65 GeV.}
Both the $S-$ and $P-$ wave amplitudes are Lorentz, gauge and parity invariant. For the sake of brevity we
will denote the $S-$ wave and $P-$ wave amplitudes by $S$ and $P$ respectively. 
To test our model and to make comparison with experimental data we have used the moments
of angular distribution. Definitions of these moments read:
\begin{equation}
\begin{split}
\langle Y^0_{\; 0}&\rangle=\frac{\mathcal{N}}{\sqrt{4\pi}}(|S|^2+|P_{-1}|^2+|P_{0}|^2+|P_{1}|^2),\\
\langle Y^1_{\; 0}&\rangle=\frac{\mathcal{N}}{\sqrt{4\pi}}(SP_0^{*}+S^*P_0), \\
\langle Y^1_{\; 1}&\rangle=\frac{\mathcal{N}}{\sqrt{4\pi}}(P_1S^*-SP_{-1}^*),\\
\langle Y^2_{\; 0}&\rangle=\frac{\mathcal{N}}{\sqrt{4\pi}}\sqrt{\frac{1}{5}}(2|P_0|^2-|P_1|^2-|P_{-1}|^2),\\
\langle Y^2_{\; 1}&\rangle=\frac{\mathcal{N}}{\sqrt{4\pi}}\sqrt{\frac{3}{5}}(P_1P_0^*-P_0P_{-1}^*),\\
\langle Y^2_{\; 2}&\rangle=\frac{\mathcal{N}}{\sqrt{4\pi}}\sqrt{\frac{6}{5}}(-P_1P_{-1}^*),
\end{split}
\end{equation}
where $\mathcal{N}$ is the normalisation factor. In formulas (8) the summation over the photon and proton helicities is implicit.
\section{Numerical calculations}
We have introduced the complex parameters multiplying the $S-$ wave and $P_0$ amplitudes to account for some phenomenological effects.
The background present in the $K^+K^-$ mass distribution and moments was parameterised using linear functions of $M_{KK}$ thus adding new
parameters. The total number of the model parameters to be fitted was 9 for the Daresbury data and 8 for the DESY data.
Results of our numerical calculations for the incident photon energies of 4 GeV and 5.65 GeV are shown in
Figs. 4 and 5, respectively. 
\begin{figure} \label{4GeV}
\begin{center}
\includegraphics[width=0.9\textwidth]{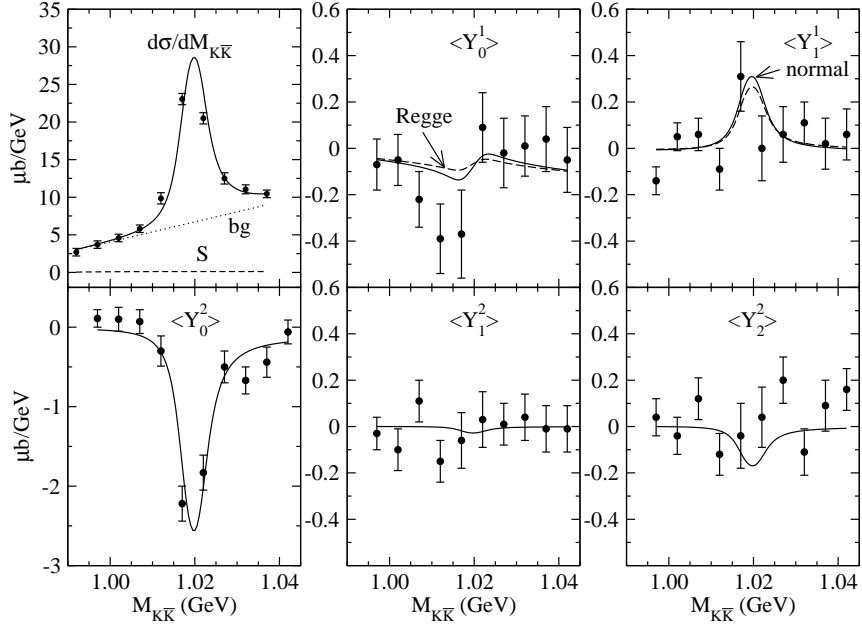}\\
\caption{Effective mass distribution and moments at $E_\gamma$=4 GeV. The experimental data are from \cite{Barber}. The curves marked by S and bg denote the $S-$ wave cross section and the background, respectively.}
\end{center}
\end{figure}
\begin{figure} \label{565GeV}
\begin{center}
\includegraphics[width=0.9\textwidth]{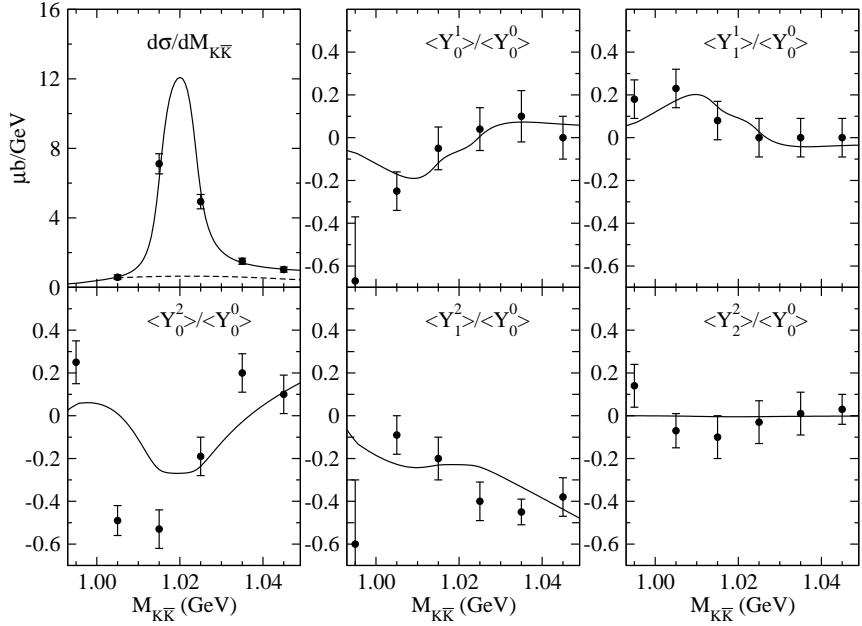}\\
\caption{Effective mass distribution and moments at $E_\gamma$=5.65 GeV. The experimental data are from \cite{Behrend}.}
\end{center}
\end{figure} 
In these figures one can see a very good agreement of the model
with experimental data.
The values of cross sections, expressed in nanobarns and computed using phenomenological parameters obtained in the minimisation procedure, are shown in Table 1. 

\begin{table}[h]
\caption{Integrated cross sections in nb}
\label{summary}
\begin{center}
\begin{tabular}{ccccc}
\hline
photon energy&\multicolumn{2}{c}{4 GeV}&\multicolumn{2}{c}{5.65 GeV}\\
\hline
$S$-wave propagator&normal&Regge&normal&Regge\\
\hline
Sum of $P$-waves&\multicolumn{2}{c}{$218.4\pm 39.5$}&\multicolumn{2}{c}{$120.5\pm 9.4$}\\
$P_{0}$-wave&$6.4_{-4.8}^{+5.5}$&$4.7_{-4.5}^{+5.7}$&$13.8_{-4.7}^{+5.3}$&$14.0_{-4.8}^{+5.3}$\\
$S$-wave&$4.9_{-3.6}^{+5.8}$&$4.3_{-3.6}^{+6.6}$&$7.0_{-4.4}^{+6.8}$&$6.8_{-4.3}^{+6.6}$\\
background&$299.4_{-10.4}^{+10.0}$&$300.0_{-10.7}^{+10.0}$&$4.5_{-6.1}^{+4.3}$&$4.7_{-5.8}^{+4.2}$\\
$|t|_{max}$&\multicolumn{2}{c}{1.5 GeV$^2$}&\multicolumn{2}{c}{0.2 GeV$^2$}\\
$M_{KK}$ range&\multicolumn{2}{c}{(0.997,1.042) GeV}&
\multicolumn{2}{c}{(1.01,1.03) GeV}\\
\hline
\end{tabular}
\end{center}
\end{table}
The most interesting result of this calculation is the value of $S-$ wave total cross section. 
Using the normal propagators its value varies from 4.9 to 7 nb for two analysed photon energies. For the Regge
propagators the values are quite similar.
This strongly supports the estimation of the $S-$ wave photoproduction cross section made by
the DESY group of Behrend et al.

We have also applied the model constructed to compute the mass distribution and the moments of the angular distribution for the incident photon energy of
$E_\gamma$=8 GeV which is the energy designed for the upgraded JLab accelerator facility \cite{Dzierba}. Results of these calculations are shown in Fig. 6.
\begin{figure} \label{Fig5}
\begin{center}
\includegraphics[width=0.9\textwidth]{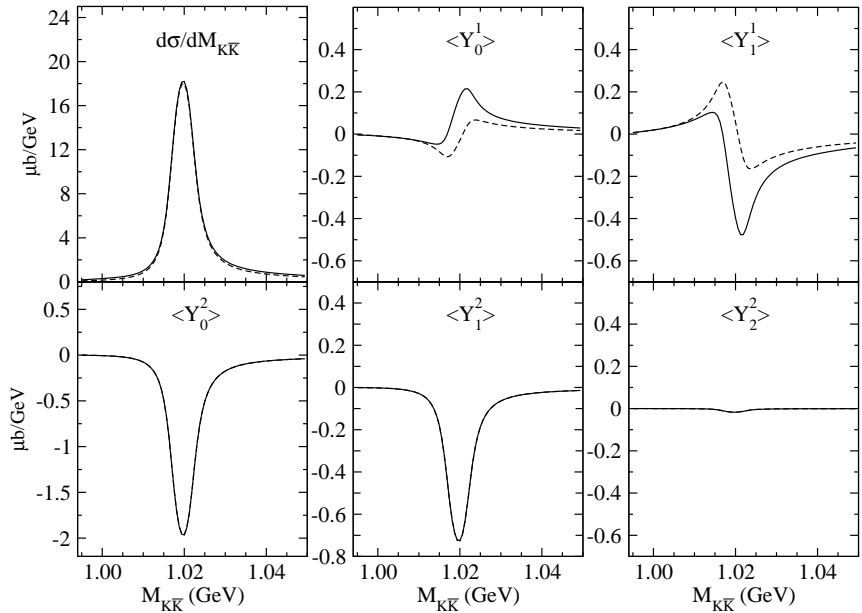}
\caption{Prediction for the mass distribution and moments at $E_\gamma$=8 GeV}
\end{center}
\end{figure}
\section{Summary and outlook}
We have shown that the $S-$ wave contribution to the elastic $K^+K^-$ photoproduction gives
a measurable effect. Our model supports the lower estimation of the $S-$ wave total photoproduction cross section
with the values between 4.9 and 7 nb. The natural consequence of these studies is an examination of the other
production reactions where partial wave interference may take place. The $\pi^+\pi^-$ photoproduction (or
electroproduction) on hydrogen is an obvious choice. One may expect an appearance  of the interference effects from the
$\rho$-dominated $P-$ wave and from the $f_0(980)$ resonance in the $S-$ wave. The recent results obtained by the HERMES
collaboration \cite{HERMES} which indicate a possible contribution from the $f_0(980)$ resonance in the $S-$ wave and $f_2(1270)$ in the $D-$ wave make this investigation even more interesting.

\end{document}